\documentclass[12pt,reqno]{amsart}
\usepackage{amsfonts,color}
\usepackage{mathrsfs}
\usepackage{graphicx} 
\usepackage{epsfig}
\usepackage{amsmath, amsthm}
\usepackage{amssymb}
\usepackage{subfigure,psfrag}
\usepackage[capposition=top]{floatrow}
\usepackage[left=2.5cm, right=2.5cm, top=2.5cm, bottom=2.5cm]{geometry}

\numberwithin{equation}{section}
\date{}

\theoremstyle{plain}
\newtheorem{theorem}{Theorem}[section]
\newtheorem{proposition}[theorem]{Proposition}
\newtheorem{lemma}[theorem]{Lemma}
\newtheorem{corollary}[theorem]{Corollary}
\newtheorem{remark}{Remark}[section]
\theoremstyle{definition}
\newtheorem{definition}[theorem]{Definition}
\newtheorem{example}[theorem]{Example}
\numberwithin{equation}{section}

\begin{document}

\def\calL{\mathcal{L}}
\def\calG{\mathcal{G}}
\def\calD{\mathcal{D}}
\def\calJ{\mathcal{J}}
\def\calM{\mathcal{M}}
\def\calN{\mathcal{N}}
\def\calO{\mathcal{O}}
\def\calA{\mathcal{A}}
\def\calS{\mathcal{S}}
\def\calP{\mathcal{P}}
\def\calU{\mathcal{U}}
\def\calK{\mathcal{K}}
\def\frakgl{\mathfrak{gl}}
\def\frako{\mathfrak{o}}
\def\fraku{\mathfrak{u}}
\def\frakg{\mathfrak{g}}
\def\frakso{\mathfrak{so}}
\def\fraksl{\mathfrak{sl}}
\def\fraksp{\mathfrak{sp}}
\def\fraksu{\mathfrak{su}}
\def\F{\mathbb{F}}
\def\R{\mathbb{R}}
\def\N{\mathbb{N}}
\def\C{\mathbb{C}}
\def\M{\mathbb{M}}
\def\H{\mathbb{H}}
\def\P{\mathbb{P}}
\def\al{\alpha}
\def\be{\beta}
\def\p{\partial}
\def\n{\, | \, }
\def\ti{\tilde}
\def\a{\alpha}
\def\r{\rho}
\def\l{\lambda}
\def\hcalG {\hat{\mathcal{G}}}
\def\diag{{\rm diag \/ }}
\def\det{{\rm det \/ }}
\def\sp{{\rm span \/ }}
\def\rd{{\rm d\/}}
\def\K{\nabla}
\def\g{\gamma}
\def\Re{{\rm Re\/}}
\def\a{\alpha}
\def\b{\beta}
\def\d{\delta}
\def\D{\triangle}
\def\e{\epsilon}
\def\g{\gamma}
\def\G{\Gamma}
\def\K{\nabla}
\def\l{\lambda}
\def\L{\Lambda}
\def\n{\,\vert\,}
\def\o{\theta}
\def\w{\omega}
\def\W{\Omega}
\def\ca{{\mathcal{A}}}
\def\cd{{\mathcal{D}}}
\def\cf{{\mathcal{F}}}
\def\cg{{\mathcal{G}}}
\def\ch{{\mathcal{H}}}
\def\ck{{\mathcal{K}}}
\def\cl{{\mathcal{L}}}
\def\cL{{\mathcal{L}}}
\def\cm{{\mathcal{M}}}
\def\cn{{\mathcal{N}}}
\def\co{{\mathcal{O}}}
\def\cp{{\mathcal{P}}}
\def\cs{{\mathcal{S}}}
\def\ct{{\mathcal{T}}}
\def\cu{{\mathcal{U}}}
\def\cv{{\mathcal{V}}}
\def\cx{{\mathcal{X}}}
\def\li{\langle}
\def\ri{\rangle}
\def\n{\ \vert\ }
\def\tr{{\rm tr}}
\def\bs{\bigskip}
\def\ms{\medskip}
\def\ss{\smallskip}
\def\hb{\hfil\break\vskip -12pt}

\def\di{$\diamond$}
\def\ni{\noindent}
\def\ti{\tilde}
\def\p{\partial}
\def\Re{{\rm Re\/}}
\def\Im{{\rm Im\/}}
\def\I{{\rm I\/}}
\def\II{{\rm II\/}}
\def\diag{{\rm diag}}
\def\ad{{\rm ad}}
\def\Ad{{\rm Ad}}
\def\Iso{{\rm Iso}}
\def\Gr{{\rm Gr}}
\def\sgn{{\rm sgn}}

\def\rd{{\rm d\/}}

\def\R{\mathbb{R} }
\def\C{\mathbb{C}}
\def\H{\mathbb{H}}
\def\N{\mathbb{N}}
\def\Z{\mathbb{Z}}
\def\O{\mathbb{O}}
\def\F{\mathbb{F}}

\def\fg{\mathfrak{G}}

\newcommand{\beg}{\begin{example}}
\newcommand{\eeg}{\end{example}}
\newcommand{\bthm}{\begin{theorem}}
\newcommand{\ethm}{\end{theorem}}
\newcommand{\bprop}{\begin{proposition}}
\newcommand{\eprop}{\end{proposition}}
\newcommand{\bcor}{\begin{corollary}}
\newcommand{\ecor}{\end{corollary}}
\newcommand{\blem}{\begin{lemma}}
\newcommand{\elem}{\end{lemma}}
\newcommand{\bca}{\begin{cases}}
\newcommand{\eca}{\end{cases}}
\newcommand{\brem}{\begin{remark}}
\newcommand{\erem}{\end{remark}}
\newcommand{\bpm}{\begin{pmatrix}}
\newcommand{\epm}{\end{pmatrix}}
\newcommand{\bbm}{\begin{bmatrix}}
\newcommand{\ebm}{\end{bmatrix}}
\newcommand{\bvm}{\begin{vmatrix}}
\newcommand{\evm}{\end{vmatrix}}
\newcommand{\bdefn}{\begin{definition}}
\newcommand{\edefn}{\end{definition}}
\newcommand{\bsub}{\begin{subtitle}}
\newcommand{\esub}{\end{subtitle}}
\newcommand{\bex}{\begin{example}}
\newcommand{\eex}{\end{example}}
\newcommand{\ben}{\begin{enumerate}}
\newcommand{\een}{\end{enumerate}}
\newcommand{\bpf}{\begin{proof}}
\newcommand{\epf}{\end{proof}}

\newcommand{\balign}{\begin{align}}
\newcommand{\ealign}{\end{align}}
\newcommand{\baligns}{\begin{align*}}
\newcommand{\ealigns}{\end{align*}}
\newcommand{\beq}{\begin{equation}}
\newcommand{\eeq}{\end{equation}}
\newcommand{\beqs}{\begin{equation*}}
\newcommand{\eeqs}{\end{equation*}}
\newcommand{\beqa}{\begin{eqnarray}}
\newcommand{\eeqa}{\end{eqnarray}}
\newcommand{\beqas}{\begin{eqnarray*}}
\newcommand{\eeqas}{\end{eqnarray*}}

\def\pdo{$\psi$do}

\def\calA{{\mathcal A}}
\def\calB{{\mathcal B}}
\def\calD{{\mathcal D}}
\def\calF{{\mathcal F}}
\def\calG{{\mathcal G}}
\def\calJ{{\mathcal J}}
\def\calK{{\mathcal K}}
\def\calL{{\mathcal L}}
\def\calM{{\mathcal M}}
\def\calN{{\mathcal N}}
\def\calO{{\mathcal O}}
\def\calP{{\mathcal P}}
\def\calR{{\mathcal R}}
\def\calS{{\mathcal S}}
\def\calU{{\mathcal U}}
\def\calV{{\mathcal V}}

\def\li{\langle}
\def\ri{\rangle}

\def\frakP{{\mathfrak{P}}}

\def\half{\frac{1}{2}}
\def\Tr{{\rm Tr\/}}
\def\nkdv{$n\times n$ KdV}

\def \a {\alpha}
\def \b {\beta}
\def \d {\delta}
\def \D {\triangle}
\def \e {\epsilon}
\def \g {\gamma}
\def \G {\Gamma}
\def \K {\nabla}
\def \l {\lambda}
\def \L {\Lambda}
\def \n {\,\vert\,}
\def \N {\,\Vert\,}
\def \o {\theta}
\def\w{\omega}
\def\W{\Omega}
\def \s {\sigma}
\def \S {\Sigma}

\def\ca{{\mathcal {A}}}
\def\cC{{\mathcal {C}}}
\def\cg{{\mathcal {G}}}
\def\ci{{\mathcal {I}}}
\def\ck{{\mathcal {K}}}
\def\cl{{\mathcal {L}}}
\def\cm{{\mathcal {M}}}
\def\cn{{\mathcal {N}}}
\def\co{{\mathcal {O}}}
\def\cp{{\mathcal {P}}}
\def\cs{{\mathcal {S}}}
\def\ct{{\mathcal {T}}}
\def\cu{{\mathcal {U}}}
\def\ch{{\mathcal {H}}}

\def\R{{\mathbb{R}}}
\def\C{{\mathbb{C}}}
\def\H{{\mathbb{H}}}
\def\Z{{\mathbb{Z}}}

\def\Re{{\rm Re\/}}
\def\Im{{\rm Im\/}}
\def\tr{{\rm tr\/}}
\def\Id{{\rm Id\/}}
\def\I{{\rm I\/}}
\def\II{{\rm II\/}}
\def\li{\leftrangle}
\def\ri{rightrangle}
\def\id{{\rm Id}}
\def\gk{\frac{G}{K}}
\def\uk{\frac{U}{K}}

\def\p{\partial}
\def\li{\langle}
\def\ri{\rangle}
\def\ti{\tilde}
\def\i{\/ \rm i }
\def\j{\/ \rm j }
\def\k{\/ \rm k}
\def\n {\ \vert\ }
\def\bu{$\bullet$}
\def\ni{\noindent}
\def\ii{{\rm i\,}}

\def\bs{\bigskip}
\def\ms{\medskip}
\def\ss{\smallskip}

\title[]{A $q$-generalization of the Toda equations for the $q$-Laguerre/Hermite Orthogonal Polynomials}
 
\author{Chuan-Tsung Chan$^\dagger$ \and Hsiao-Fan Liu$^{\ddagger}$}
\address{}
\dedicatory{$^\dagger$ Department of Applied Physics, Tunghai University\\
$^{\ddagger}$Department of Mathematics, National Tsing Hua University\\
$^\dagger$ ctchan@go.thu.edu.tw, $^{\ddagger}$ hfliu@math.nthu.edu.tw}

\date{\today} 

\begin{abstract}


Based on the motivation of generalizing the correspondence between the Lax equation for the Toda lattice and the deformation theory of the orthogonal polynomials, we derive a $q$-deformed version of the Toda 
equations for both $q$-Laguerre/Hermite ensembles, and check the compatibility 
with the quadratic relation.

\end{abstract}

\maketitle

\lineskip=0.25cm
\section{Introduction}\

There is an interesting correspondence between the Lax pair formulation of the Toda system \cite{F74,HF74} and the deformation theory of the orthogonal polynomial systems \cite{C13}. In the simplest scenario, Flaschka's Lax pair equation \cite{Lax68} for the one-dimensional periodic Toda lattice \cite{MT67,MT70}
\beqa
H&:=&\sum_n \frac{1}{2} P_n^2+\left(e^{Q_{n-1}-Q_n}-1\right),\\
a_n&:=&e^{(Q_{n-1}-Q_n)/2},\quad a_0:=a_N,\\
b_n&:=&-P_n, \quad b_0:=b_N,
\eeqa

\beq
L:=\left(\begin{array}{ccccc}b_0 & a_1 & 0 & \cdots & a_N \\a_1 & b_1 & a_2 & \cdots & 0 \\0 & a_2 & b_2 & \cdots & 0 \\ \vdots & \vdots & \vdots & \vdots & \vdots \\a_N & 0 & 0 & 0 & b_{N-1}\end{array}\right)=L^t,
\eeq
\beq
B:=\left(\begin{array}{ccccc}0 & a_1 & 0 & \cdots & -a_N \\-a_1 & 0 & a_2 & \cdots & 0 \\0 & -a_2 & 0 & \cdots & 0 \\ \vdots & \vdots & \vdots & \vdots & \vdots \\ a_N & 0 & 0 & -a_{N-1} & 0\end{array}\right)=-B^t,
\eeq
\beq
2\dot L=[B,L]=BL-LB,
\eeq
is equivalent to the Hamiltonian equations of the Toda system,
\beqa
\dot a_n&=&\frac{a_n}{2}\left(b_{n}-b_{n-1}\right) \Leftrightarrow \dot Q_n=\frac{\p H}{\p P_n}=P_n,\label{intro1}\\
\dot b_n&=&\left(a_{n+1}^2-a_{n}^2\right) \Leftrightarrow \dot P_n=-\frac{\p H}{\p Q_n}=e^{Q_{n-1}-Q_n}-e^{Q_n-Q_{n+1}}.\label{intro2}
\eeqa

On the other hand, for general orthonormal polynomial system with a weight function $w_0(x)$,
\beq
\int p_m(x)p_n(x) w_0(x)~dx=\d_{mn},
\eeq
there exist the well-known three-term recursive relations \cite{IS11,W07,BW10},
\beq
xp_n(x)=a_{n+1}p_{n+1}(x)+b_np_n(x)+a_np_{n-1}(x).
\eeq
Upon the following exponential deformation (or time evolution)
\beq
w_0(x) \rightarrow w(x;t):=w_0(x)e^{-tx},
\eeq
the orthogonal polynomials and the recursive coefficients acquire $t$-dependences:
\beq
p_n(x) \rightarrow p_n(x,t), \quad a_n \rightarrow a_n(t), \quad b_n \rightarrow b_n(t),
\eeq
and one can show that they satisfy the same relation, Eqs.\eqref{intro1},\eqref{intro2} (see Theorem \ref{thm2-1} below).

In addition to the correspondence mentioned above, given two sets of orthogonal polynomials, if their weights are related by a quadratic relation
\beq
w(x;t)=|x|v(x^2;t),
\eeq 
then we can show that the recursive coefficient of the former set of orthogonal polynomials can be related to that of the later (The details will be given in Sec. $2$). 

In this paper, we wish to generalized the correspondence and check the compatibility of quadratic relation in a discrete setting of the orthogonal polynomials. For this purpose, we focus on a q-generalization of the Laguerre polynomials and the corresponding $q$-Hermite polynomials. Our main goal is to derive the corresponding $q$ deformation/evolution equations for each systems.

This paper is organized as follows: to set up the notations and introduce the idea of quadratic relation, we give a brief review of $q$-Laguerre and $q$-Hermite polynomials in Sec. $2$. Then we derive the $q$-Toda equations for the Laguerre/Hermite polynomials in Sec. $3$. A brief summary and conclusion will be addressed in Sec. $4$.

\section{The Quadratic Relations for the Laguerre/Hermite Orthogonal Polynomials}

\subsection{Review of the quadratic relation among recursive coefficients for the orthogonal polynomials associated with the classical 
             Laguerre/Hermite weights}\label{sec:1}\
\vspace{0.5cm}   



Given the classical Laguerre weight defined as 
\beq\label{}
v^{(\a)}(x;\kappa):= x^\a \exp(-\kappa x),\quad 0 \leq \kappa, \quad -1<\a, \quad 0 \leq x,
\eeq
we can compute the orthonormal polynomials $p_n^{(\a)}(x;\kappa)$ as
\beq\label{jc}
\int_0^\infty p_m^{(\a)}(x;\kappa)p_n^{(\a)}(x;\kappa) v^{(\a)}(x;\kappa) dx =\d_{mn} 
\eeq
through Gram-Schmidt process.

Similarly, from the classical Hermite weight, 
\beq\label{}
\w^{(\a)}(x;\kappa):=|x|^{2\a+1}\exp(-\kappa^2x^2), \quad 0 \leq \kappa, \quad x \in \R,
\eeq 
we obtain associated orthonormal polynomials $P_n^{(\a)}(x,\kappa)$ as
\beq\label{}
\int_{-\infty}^\infty P_m^{(\a)}(x;\kappa)P_n^{(\a)}(x;\kappa)\w^{(\a)}(x;\kappa) dx=\d_{mn}.
\eeq

The quadratic relation among these two sets of orthonormal polynomials
is based on a simple connection between the Lagurre and Hermite 
weights. Namely, 
\beq\label{ba}
\w^{(\a)}(x;\kappa)=|x|v^{(\a)}(x^2;\kappa^2).
\eeq 
One immediate consequence of Eq. \eqref{ba} is that, the orthonormal 
polynomials $P_n^{(\a)}(x;\kappa)$ can be expressed in 
terms of the orthogonal polynomials $p_n^{(\a)}(x;\kappa)$ as follows,
\beq\label{}
P_{2n}^{(\a)}(x;\kappa)=p_n^{(\a)}(x^2;\kappa^2), \quad
P_{2n+1}^{(\a)}(x;\kappa)=x p_n^{(\a+1)}(x^2;\kappa^2).
\eeq

The set of orthonormal polynomials associated with any weight function can
be viewed as a complete set of basis for the function space. Hence, it
induces a natural realization of the Heisenberg algebra, $[\frac{d}{dx},x]=1.$
In particular, the matrix elements of the position operator consist of the 
three-term recursive coefficients among orthonormal polynomials. In 
the case of the Laguerre weight, it is given as
\beq\label{}
x p_n^{(\a)}(x;\kappa)=a_{n+1}^{(\a)}(\kappa) p_{n+1}^{(\a)}(x;\kappa)+b_n^{(\a)}(\kappa) p_n^{(\a)}(x;\kappa)
+a_n^{(\a)}(\kappa) p_{n-1}^{(\a)}(x;\kappa),
\eeq 
and in the case of the Hermite weight, we have
\beq\label{bb}
x P_n^{(\a)}(x;\kappa)=A_{n+1}^{(\a)}(\kappa) P_{n+1}^{(\a)}(x;\kappa)+A_n^{(\a)}(\kappa) P_{n-1}^{(\a)}(x;\kappa).
\eeq 
By computing $x^2P_n^{(\a)}(x,\kappa)$ in two ways (see Theorem \ref{thm:2-1} for details), we obtain the quadratic 
relation among the two sets of recursive coefficients:
\begin{align}
a_n^{(\a)}(\kappa^2)&=A_{2n}^{(\a)}(\kappa)A_{2n-1}^{(\a)}(\kappa), \label{2-1-1}\\
b_n^{(\a)}(\kappa^2)&=\left(A_{2n+1}^{(\a)}(\kappa)\right)^2+\left(A_{2n}^{(\a)}(\kappa)\right)^2.\label{2-1-2}
\end{align}

\subsection{On the quadratic relation between generalized $q$-Laguerre/Hermite ensembles}\
\vspace{0.5cm}

In this paper, we take generalized little $q$-Laguerre and $q$-Hermite ensembles
\cite{IS11,W07,BW10,SS94,FS13,39A13} as an illustrative example of the quadratic relation. We consider the generalized little $q$-Laguerre weight ($0 \leq \kappa < \frac{1}{q}$) 
\beq 
v^{(\a)}(x;\kappa,q) := |x|^\a (q\kappa x; q)_\infty=|x|^\a \prod_{l=0}^\infty (1-q^{l+1}\kappa x),
\eeq and the generalized $q$-Hermite 
weight 
\beq
\w^{(\a)}(x;\kappa,q) = |x|^{2\a+1} (q^2\kappa^2x^2;q^2)_\infty=|x| v^{(\a)}(x^2;\kappa^2,q^2).
\eeq

Given the orthonormal polynomials of the $q$-Laguerre ensemble $p_n^{(\a)}(x;\kappa,q)$, 
we can express the orthonormal polynomials of the $q$-Hermite ensemble $P_n^{(\a)}(x;\kappa,q)$ 
as follows:
\beqa\label{ga}
P_{2n}^{(\a)}(x;\kappa,q)&=&\sqrt{\frac{1+q}{2}}p_n^{(\a)}(x^2;\kappa^2,q^2)  \mbox{ (even, } \deg=2n),\\
P_{2n+1}^{(\a)}(x;\kappa,q)&=&\sqrt{\frac{1+q}{2}} x p_n^{(\a+1)}(x^2;\kappa^2,q^2) \mbox{ (odd, } \deg=2n+1).
\eeqa

One can easily check that $P_{2n}^{(\a)}(x;\kappa,q)$ and $P_{2n+1}^{(\a)}(x;\kappa,q)$ satisfy the orthonormal 
conditions, for instance,
\begin{align}
&\int_{-1}^1 P_{2m}^{(\a)}(x;\kappa,q) P_{2n}^{(\a)}(x;\kappa,q)\w^{(\a)}(x;\kappa,q) d_q x\nonumber\\
=&2(1-q) \sum_{k=0}^\infty P_{2m}^{(\a)}(q^k;\kappa,q) P_{2n}^{(\a)}(q^k;\kappa,q)\w^{(\a)}(q^k;\kappa,q) q^k\nonumber\\
=&2(1-q)\left(\frac{1+q}{2}\right)\sum_{k=0}^\infty p_m^{(\a)}(q^{2k};\kappa^2,q^2) p_n^{(\a)}(q^{2k};\kappa^2,q^2) q^k v^{(\a)}(q^{2k};\kappa^2,q^2) q^k\nonumber\\
=&\int_0^1 p_m^{(\a)}(x;\kappa^2,q^2) p_n^{(\a)}(x;\kappa^2,q^2) v^{(\a)}(x;\kappa^2,q^2) d_q x=\d_{mn}.
\end{align}
\begin{align}
&\int_{-1}^1 P_{2m+1}^{(\a)}(x;\kappa,q) P_{2n+1}^{(\a)}(x;\kappa,q)\w^{(\a)}(x;\kappa,q) d_q x\nonumber\\
=&2(1-q) \sum_{k=0}^\infty P_{2m+1}^{(\a)}(q^k;\kappa,q) P_{2n+1}^{(\a)}(q^k;\kappa,q)\w^{(\a)}(q^k;\kappa,q) q^k\nonumber\\
=&2(1-q)\left(\frac{1+q}{2}\right)\sum_{k=0}^\infty q^k p_m^{(\a+1)}(q^{2k};\kappa^2,q^2) q^k p_n^{(\a+1)}(q^{2k};\kappa^2,q^2) q^k v^{(\a)}(q^{2k};\kappa^2,q^2) q^k\nonumber\\
=&(1-q^2)\sum_{k=0}^\infty p_m^{(\a+1)}(q^{2k};\kappa^2,q^2) p_n^{(\a+1)}(q^{2k};\kappa^2,q^2) v^{(\a+1)}(q^{2k};\kappa^2,q^2) q^{2k}\nonumber\\
=&\int_0^1 p_m^{(\a+1)}(x;\kappa^2,q^2) p_n^{(\a+1)}(x;\kappa^2,q^2) v^{(\a+1)}(x;\kappa^2,q^2) d_q x=\d_{mn}.
\end{align}
Note that the $q$-integral (or Jackson integral) is defined in Eq.\eqref{app:3}. The $P_{2m}^{(\a)}$-$P_{2m+1}^{(\a)}$ orthogonality is trivial due to the even parity of 
the generalized $q$-Hermite weight.

Similar to the classical cases Eqs. \eqref{2-1-1}, \eqref{2-1-2}, there exists a correspondence between 
recursive coefficients associated with the generalized little $q$-Laguerre and $q$-Hermite ensembles. 
\bthm\label{thm:2-1}
The recursive coefficients associated with $q$-generalized Laguerre and Hermite ensembles satisfying 
the following relations:
\beqa
a_n^{(\a)}(\kappa^2,q^2)&=&A_{2n}^{(\a)}(\kappa,q)A_{2n-1}^{(\a)}(\kappa,q), \label{bc} \\
b_n^{(\a)}(\kappa^2,q^2)&=&\left(A_{2n+1}^{(\a)}(\kappa,q)\right)^2+\left(A_{2n}^{(\a)}(\kappa,q)\right)^2, \label{bd}\\
a_n^{(\a+1)}(\kappa^2,q^2)&=&A_{2n+1}^{(\a)}(\kappa,q)A_{2n}^{(\a)}(\kappa,q),\label{be}\\
b_n^{(\a+1)}(\kappa^2,q^2)&=&\left[A_{2n+2}^{(\a)}(\kappa,q)\right]^2+\left[A_{2n+1}^{(\a)}(\kappa,q)\right]^2.\label{bf}
\eeqa
\ethm
\bpf
\begin{align*}
&x^2 P_{2n}^{(\a)}(x;\kappa,q)\\
&=x\left[A_{2n+1}^{(\a)}(\kappa,q)P_{2n+1}^{(\a)}(x;\kappa,q)+A_{2n}^{(\a)}(\kappa,q)P_{2n-1}^{(\a)}(x;\kappa,q)\right]\\
                                   &=A_{2n+1}^{(\a)}(\kappa,q)\left[A_{2n+2}^{(\a)}(\kappa,q)P_{2n+2}^{(\a)}(x;\kappa,q)+A_{2n+1}^{(\a)}(\kappa,q)P_{2n-1}^{(\a)}(x;\kappa,q)\right]\\
                                   &+A_{2n}^{(\a)}(\kappa,q)\left[A_{2n}^{(\a)}(\kappa,q)P_{2n}^{(\a)}(x;\kappa,q)+A_{2n-1}^{(\a)}(\kappa,q)P_{2n-2}^{(\a)}(x;\kappa,q)\right]\\
                                   &=\left[A_{2n+1}^{(\a)}(\kappa,q)A_{2n+2}^{(\a)}(\kappa,q)\right]P_{2n+2}^{(\a)}(x;\kappa,q)+\left[\left(A_{2n+1}^{(\a)}(\kappa,q)\right)^2+\left(A_{2n}^{(\a)}(\kappa,q)\right)^2\right]P_{2n}^{(\a)}(x;\kappa,q)\\
                                   &+[A_{2n-1}^{(\a)}(\kappa,q)A_{2n}^{(\a)}(\kappa,q)]P_{2n-2}^{(\a)}(x;\kappa,q).
\end{align*}
On the other hand, using the expression of Eq. \eqref{ga}, we have
\begin{align*}
&x^2 P_{2n}^{(\a)}(x;\kappa,q)\\
&=x^2 \sqrt{\frac{1+q}{2}}p_n^{(\a)}(x^2;\kappa^2,q^2)\\
&=\sqrt{\frac{1+q}{2}}\left[a_{n+1}^{(\a)}(\kappa^2,q^2)p_{n+1}^{(\a)}(x^2;\kappa^2,q^2)+b_n^{(\a)}(\kappa^2,q^2)p_n^{(\a)}(x^2;\kappa^2,q^2)+a_n^{(\a)}(\kappa^2,q^2)p_{n-1}^{(\a)}(x^2;\kappa^2,q^2)\right]\\
&=a_{n+1}^{(\a)}(\kappa^2,q^2)P_{2n+2}^{(\a)}(x;\kappa,q)+b_n^{(\a)}(\kappa^2,q^2)P_{2n}^{(\a)}(x;\kappa,q)+a_n^{(\a)}(\kappa^2,q^2)P_{2n-2}^{(\a)}(x;\kappa,q).
\end{align*}
By comparing the coefficients on both expressions, we get Eqs.\eqref{bc}, \eqref{bd}.

If we examine similar calculations for the odd $q$-Hermite orthonormal 
polynomials, we get
\begin{align*}
x^2 P_{2n+1}^{(\a)}&=[A_{2n+3}^{(\a)}A_{2n+2}^{(\a)}]P_{2n+3}^{(\a)}\\
&+[\left(A_{2n+2}^{(\a)}\right)^2+\left(A_{2n+1}^{(\a)}\right)^2]P_{2n+1}^{(\a)}+[A_{2n+1}^{(\a)}A_{2n}^{(\a)}]P_{2n-1}^{(\a)}.
\end{align*}
Alternatively,
\begin{align*}
&x^2 P_{2n+1}^{(\a)}(x;\kappa,q)\\
&=x^2 \sqrt{\frac{1+q}{2}}x p_n^{(\a+1)}(x^2;\kappa^2,q^2)\\
&=\sqrt{\frac{1+q}{2}}\left[a_{n+1}^{(\a+1)}(\kappa^2,q^2)p_{n+1}^{(\a+1)}(x^2;\kappa^2,q^2)+b_n^{(\a+1)}(\kappa^2,q^2)p_n^{(\a+1)}(x^2;\kappa^2,q^2)+a_n^{(\a+1)}(\kappa^2,q^2)p_{n-1}^{(\a+1)}(x^2;\kappa^2,q^2)\right]\\
&=a_{n+1}^{(\a+1)}(\kappa^2,q^2)P_{2n+3}^{(\a)}(x;\kappa,q)+b_n^{(\a+1)}(\kappa^2,q^2)P_{2n+1}^{(\a)}(x;\kappa,q)+a_n^{(\a+1)}(\kappa^2,q^2)P_{2n-1}^{(\a)}(x;\kappa,q).
\end{align*}
Thus, we have shown Eqs.\eqref{be}, \eqref{bf}.
\epf
Eliminating the recursive coefficients of the $q$-Laguerre orthonormal polynomials, 
$a_n^{(\a)},b_n^{(\a)}$, in both sets of the equation, we obtain
\beq\label{}
A_{2n}^{(\a)}(\kappa,q)A_{2n-2}^{(\a)}(\kappa,q)=A_{2n+1}^{(\a-1)}(\kappa,q)A_{2n}^{(\a-1)}(\kappa,q),
\eeq
and
\beq\label{}
\left(A_{2n+1}^{(\a)}(\kappa,q)\right)^2+\left(A_{2n}^{(\a)}(\kappa,q)\right)^2=\left(A_{2n+2}^{(\a-1)}(\kappa,q)\right)^2+\left(A_{2n+1}^{(\a-1)}(\kappa,q)\right)^2.
\eeq
For the general $\kappa$ case, we check the compatibility between the quadratic relation 
and the evolution equations (w.r.t $\kappa$) in Sec. $3$.

\subsection{On the compatibility of the quadratic relation with deformation}\
\vspace{0.5cm}

As mentioned in the introduction, our main calculations are about the derivation of the discrete evolutions of the recursive coefficients with respect to parameter $\kappa$. Before we present the details, it is useful to recall the basic idea in the classical case.

To begin with, we shall study the differential equations associated with the recursive coefficients of the Laguerre/Hermite polynomials under the deformation of the weight.

\bthm\label{thm2-1}
For the weight function of the Laguerre ensembles,
\beq
v^{(\a)}(x;t):=x^\a \exp(-tx),
\eeq
the recursive coefficients $a^{(\a)}_n(t), b^{(\a)}_n(t)$, defined as
\beq
xp^{(\a)}_n(x;t)=a^{(\a)}_{n+1}(t)p^{(\a)}_{n+1}(x;t)+b^{(\a)}_n(t)p^{(\a)}_n(x;t)+a^{(\a)}_n(t)p^{(\a)}_{n-1}(x;t),
\eeq
satisfy the following differential equations
\beqa\label{sec_2-3:2}
\dot a_n^{(\a)}&=&\frac{a_n^{(\a)}}{2}\left(b^{(\a)}_n-b^{(\a)}_{n-1}\right),\nonumber\\
\dot b_n^{(\a)}&=&\left(a_{n+1}^{(\a)}\right)^2-\left(a_n^{(\a)}\right)^2.
\eeqa
\ethm 
In order to compute the time derivatives of the recursive coefficients $a_n^{(\a)},b_n^{(\a)}$, we first compute the Fourier expansion of the time derivatives of the orthogonal polynomials:
\beq\label{sec_2-3:1}
\frac{\p}{\p{t}}p_n^{(\a)}(x;t)=\sum_{k=0}^n p_k^{(\a)}(x;t) C_{k n}^{(\a)}.
\eeq
\bthm\label{sec2:3-4}
\beq
\frac{\p}{\p{t}}p_n^{(\a)}(x;t)=\frac{b_n^{(\a)}}{2}p_n^{(\a)}(x;t)+a_n^{(\a)}p_{n-1}^{(\a)}(x;t).
\eeq
\ethm
\bpf
By making the following projections, we can prove that the only non-zero terms in the expansion \eqref{sec_2-3:1} are $C^{(\a)}_{n-1,n}=a_n^{(\a)}$ and $C^{(\a)}_{nn}=\frac{b_n^{(\a)}}{2}$. For $C^{(\a)}_{nn}$,
\beqa
&&\frac{d}{dt} \int p_n^{(\a)}p_n^{(\a)}v^{(\a)}~dx=0\\
&\Rightarrow& 2 \int p_n^{(\a)}\left[\frac{\p}{\p t}p_n^{(\a)}\right]v^{(\a)}~dx+\int \left[p_n^{(\a)}\right]^2\left[\frac{\p}{\p t} v^{(\a)}\right]~dx=0.
\eeqa
Note that 
\beq
\frac{\p}{\p t} v^{(\a)}(x;t)=-xv^{(\a)}(x;t),
\eeq
and the first term gives $2C_{nn}^{(\a)}$. So we conclude that $C_{nn}^{(\a)}=\frac{1}{2}b_n^{(\a)}$.

Similarly, for $C^{(\a)}_{n-1,n}$,
 \beq
 \frac{d}{dt}\left[ \int p_n^{(\a)}p_m^{(\a)}v^{(\a)}~dx\right]=0, \quad (\mbox{for } m<n). 
 \eeq 
 If $m=n-1$, we derive $C^{(\a)}_{n-1,n}=a_n^{(\a)}$, and if $m<n-1$, we obtain $C^{(\a)}_{mn}=0$. In conclusion, we get
 \beq
\frac{\p}{\p{t}}p_n^{(\a)}(x;t)=\frac{b_n^{(\a)}}{2}p_n^{(\a)}(x;t)+a_n^{(\a)}p_{n-1}^{(\a)}(x;t).
\eeq
\epf
Having computed the Fourier coefficients of $\frac{\p}{\p{t}}p_n^{(\a)}(x;t)$, we now derive the evolution equations for $a_n^{(\a)}$ and $b_n^{(\a)}$.
\bpf[Proof of Theorem \ref{thm2-1}]
We can compute the time-evolution of $xp_n^{(\a)}(x;t)$ in two ways. Firstly, 
\beqa
\frac{\p}{\p t}\left[xp_n^{(\a)}\right]&=&\frac{\p}{\p t}\left[a^{(\a)}_{n+1}p^{(\a)}_{n+1}+b^{(\a)}_np^{(\a)}_n+a^{(\a)}_np^{(\a)}_{n-1}\right]\\
&=&\left(\dot a^{(\a)}_{n+1}+\frac{1}{2}a^{(\a)}_{n+1}b^{(\a)}_{n+1}\right) p^{(\a)}_{n+1}+\left[\left(a^{(\a)}_{n+1}\right)^2+\dot b^{(\a)}_{n}+\frac{1}{2}\left(b^{(\a)}_{n}\right)^2\right]p^{(\a)}_{n}\nonumber\\
&&+\left(a^{(\a)}_{n}b^{(\a)}_{n}+\dot a^{(\a)}_{n}+\frac{1}{2}a^{(\a)}_{n}b^{(\a)}_{n-1}\right)p^{(\a)}_{n-1}+\left(a^{(\a)}_{n}a^{(\a)}_{n-1}\right)p^{(\a)}_{n-2}.
\eeqa
Next, 
\beqa
&&\frac{\p}{\p t}\left[x p^{(\a)}_{n}\right]=x\left[\frac{\p}{\p t} p^{(\a)}_{n}\right]\nonumber\\
&=&x\left[\frac{b_n^{(\a)}}{2}p_n^{(\a)}+a_n^{(\a)}p_{n-1}^{(\a)}\right]\nonumber\\
&=&\left(\frac{a^{(\a)}_{n+1}b^{(\a)}_{n}}{2}\right)p^{(\a)}_{n+1}+\left[\left(a^{(\a)}_{n}\right)^2+\frac{1}{2}\left(b^{(\a)}_{n}\right)^2\right]p^{(\a)}_{n}+\left(a^{(\a)}_{n}b^{(\a)}_{n-1}+\frac{1}{2}a^{(\a)}_{n}b^{(\a)}_{n}\right)p^{(\a)}_{n-1}\nonumber\\
&&+\left(a^{(\a)}_{n}a^{(\a)}_{n-1}\right)p^{(\a)}_{n-2}.
\eeqa
By comparing the two equations, we prove Eq.\eqref{sec_2-3:2}.
\epf
Similar calculations can be applied to the case of the Hermite ensembles, so we simply state the results.
\bthm\label{sec2:3-5}
For Hermite ensembles of orthonormal polynomials defined by the weight function,
\beq
w^{(\a)}(x;t):=|x|^{2\a+1}\exp(-tx^2),
\eeq 
the Fourier expansion of the time-derivative of $P_n^{(\a)}$ is given as
\beq
\frac{\p}{\p{t}}P_n^{(\a)}=\frac{1}{2}\left[\left(A_{n+1}^{(\a)}\right)^2+\left(A_{n}^{(\a)}\right)^2\right]P_n^{(\a)}+\left[A_{n-1}^{(\a)}A_n^{(\a)}\right]P_{n-2}^{(\a)}.
\eeq
\ethm

\bthm
For Hermite ensembles, the recursive coefficients in the recurrence relation,
\beq
xP^{(\a)}_n(x;t)=A^{(\a)}_{n+1}(t)P^{(\a)}_{n+1}(x;t)+A^{(\a)}_n(t)p^{(\a)}_{n-1}(x;t),
\eeq
satisfy the following differential equation
\beq\label{sec_2-3:3}
\dot A_n^{(\a)}=\frac{1}{2}A_n^{(\a)}\left[\left(A^{(\a)}_{n+1}\right)^2-\left(A^{(\a)}_{n-1}\right)^2\right].
\eeq
\ethm
Note that the differential equation for the recursive coefficients, Eq.\eqref{sec_2-3:3}, is also known as the Volterra equation. See \cite{35P25,M75,58F07} for further elaborations about the Lax pair formulation of this equation.

Having obtained the evolution equations, Eqs.\eqref{sec_2-3:2}, \eqref{sec_2-3:3}, for the Laguerre and Hermite ensembles, one can check that the quadratic relations, Eqs.\eqref{2-1-1}, \eqref{2-1-2}, are compatible with the time evolutions. Our aim in this paper is to generalize these computations to a fully $q$-discretized evolution of the $q$-Laguerre/Hermite orthogonal polynomial systems.

\section{$q$-Generalization of the Toda equations from $\kappa$-deformation of the little $q$-Laguerre/Hermite ensembles}

\subsection{$q$-Difference equations for the recursive coefficients of the $q$-Laguerre
polynomials}\
\vspace{0.5cm}


In this section, we study the $q$-difference equations describing the $\kappa$ dependence 
of the recursive coefficients $a_n^{(\a)}(\kappa),b_n^{(\a)}(\kappa)$ associated with the 
generalized little $q$-Laguerre ensemble. In the classical case, such equations correspond to 
the Lax equation of the Toda equations \cite{F74,HF74}. Hence, our results provide a $q$-generalization 
of the classical Toda equation. To achieve this, we shall express the Fourier expansion (w.r.t $\kappa$ variable) of the $q$-derivative 
on the $q$-Laguerre orthonormal polynomials in terms of the recursive coefficients,
\beq\label{ia}
\calD_q^\kappa p_n^{(\a)}(x,\kappa)=\sum_{j=0}^n p_j^{(\a)}(x,\kappa) \xi_{j n}^{(\a)}(\kappa).
\eeq
Following similar discussion as Theorem \ref{sec2:3-4}, we can express the Fourier coefficients of the $\kappa$-deformation of the orthonormal polynomials in terms of the recursive coefficients, $a_n^{(\a)}$ and $b_n^{(\a)}$.

\bthm\label{thm3-1}
The Fourier coefficients of the $\kappa$-deformation of the orthonormal polynomials associated with the little $q$-Laguerre weight is given as $(\l:=(1-q)\kappa)$
\beq\label{eq3-1-1}
\calD_q^\kappa p_n^{(\a)}(x,\kappa)=p_n^{(\a)}(x,\kappa)\xi_{nn}^{(\a)}(\kappa)+p_{n-1}^{(\a)}\xi_{n-1,n}^{(\a)}(\kappa).
\eeq
Here 
\beq\label{3-1-a}
\xi_{n-1,n}^{(\a)}(\kappa)=\frac{\dfrac{q}{1-q}\bar a_n^{(\a)}}{1-\l \xi_{n-1,n-1}^{(\a)}}=\left(\frac{q}{1-q}\right)\frac{a_n^{(\a)}}{1-\l \xi_{nn}^{(\a)}},
\eeq
\beq\label{3-1-b}
\xi_{nn}^{(\a)}(\kappa)=\frac{1}{\sqrt{2}(1-q)\kappa}\left\{\sqrt{2}-\sqrt{(1-q\kappa \bar b_n^{(\a)})+\sqrt{1-2q\kappa \bar b_n^{(\a)}+4(q\kappa)^2\left[\left(\bar b_n^{(\a)}\right)^2-\left(a_n^{(\a)}\right)^2\right]}}\right\},
\eeq
and
\beq
\bar a_n^{(\a)}(\kappa):= a_n^{(\a)}(q\kappa), \quad \bar b_n^{(\a)}(\kappa):= b_n^{(\a)}(q\kappa).
\eeq
\ethm
\bpf
The main point of this theorem is to find an expression relating $\xi_{nn}^{(\a)}$ in terms of the recursive 
coefficients $a_n^{(\a)},b_n^{(\a)}$. By taking $q$-derivative w.r.t $\kappa$ variable on the orthonormal condition, and 
recalling the $q$-Leibniz rule, Eq.\eqref{app:2}, we derive a master equation among these Fourier 
coefficients.
\beq\label{}
\calD_q^\kappa \left[\int_0^1 p_m^{(\a)}(x,\kappa)p_n^{(\a)}(x,\kappa) v^{(\a)}(x,\kappa)d_q x\right]=0.
\eeq
This implies for $m\leq n$,
\beq\label{}
(1-\l \xi_{mm}^{(\a)})\xi_{mn}^{(\a)}-\l \sum_{j=0}^{m-1} \xi_{jm}^{(\a)} \xi_{jn}^{(\a)}+\d_{mn} \xi_{nn}^{(\a)}
=\d_{m,n-1}\left[\frac{q}{1-q}\bar a_n^{(\a)}\right]+\d_{mn}\left[\frac{q}{1-q}\bar b_n^{(\a)}\right].
\eeq
From these results, we can extract useful information by specifying the value of $m$:
\begin{itemize}
\item[Case $1$:] $m\leq n-2 $\\
We find, by induction, 
$\xi_{mn}^{(\a)}(\kappa)=0$, if $m \leq n-2$. Consequently, there are only two terms in the 
$q$-derivative (w.r.t $\kappa$ variable) of the $q$-Laguerre orthonormal polynomials,
\beq
\calD_q^\kappa p_n^{(\a)}(x,\kappa)=p_n^{(\a)}(x,\kappa)\xi_{nn}^{(\a)}(\kappa)+p_{n-1}^{(\a)}\xi_{n-1,n}^{(\a)}(\kappa).
\eeq
\item[Case $2$:] $m=n-1$\\
In this case, we relate the two Fourier coefficients as follows:
\beq\label{5-1-1}
\xi_{n-1,n}^{(\a)}=\frac{\dfrac{q}{1-q}\bar a_n^{(\a)}}{1-\l \xi_{n-1,n-1}^{(\a)}}=\left(\frac{q}{1-q}\right)\frac{a_n^{(\a)}}{1-\l \xi_{nn}^{(\a)}}.
\eeq
\item[Case $3$:] $m=n$\\
By suitable rearrangements, we derive a recursive equation relating the diagonal Fourier coefficients
$\xi_{nn}^{(\a)}$ to the recursive coefficients $a_n^{(\a)},b_n^{(\a)}$ as follows:
\beq\label{if}
(1-\l \xi_{nn}^{(\a)})^2+\frac{(q \kappa \bar a_n^{(\a)})^2}{(1-\l \xi_{n-1,n-1}^{(\a)})^2}=1-q \kappa \bar b_n^{(\a)}.
\eeq
\end{itemize}

By using Eq.\eqref{5-1-2}, we can rewrite Eq.\eqref{if} as a quadratic equation for $(1-\l \xi_{nn}^{(\a)})^2$,
\beq\label{5-1-3}
(1-\l \xi_{nn}^{(\a)})^2+\frac{(q \kappa a_n^{(\a)})^2}{(1-\l \xi_{nn}^{(\a)})^2}=1-q \kappa \bar b_n^{(\a)}.
\eeq
From the solution of this equation, we then obtain an expression of $\xi_{nn}^{(\a)}$ in terms of $a_n^{(\a)}$ and 
$\bar b_n^{(\a)}$,
\beq\label{5-1-4}
\xi_{nn}^{(\a)}=\frac{1}{\sqrt{2}(1-q)\kappa}\left\{\sqrt{2}-\sqrt{(1-q\kappa \bar b_n^{(\a)})+\sqrt{1-2q\kappa \bar b_n^{(\a)}+4(q\kappa)^2\left[\left(\bar b_n^{(\a)}\right)^2-\left(a_n^{(\a)}\right)^2\right]}}\right\}.
\eeq
\epf

Recalling the definition of the $q$-derivative, Eq.\eqref{app:1}, we can also transform this expansion 
formula, Eq.\eqref{eq3-1-1}, as a $q$-shifting relation ($\l:=(1-q)\kappa$):
\beq\label{}
p_n^{(\a)}(x,q\kappa)= p_n^{(\a)}(x,\kappa)[1-\l \xi_{nn}^{(\a)}(\kappa)]-\l \sum_{j=0}^{n-1} p_j^{(\a)}(x,\kappa)\xi_{j n}^{(\a)}(\kappa).
\eeq

\bthm\label{thm3-2}
The $\kappa$-deformation of the recursive coefficients associated with the generalized little $q$-Laguerre orthonormal polynomials is given by 
\beqa
\calD_q^\kappa a_n^{(\a)}(\kappa)&=&\frac{\xi_{n-1,n-1}^{(\a)}(\kappa)-\xi_{nn}^{(\a)}(\kappa)}{1-\l \xi_{nn}^{(\a)}(\kappa)} a_n^{(\a)}(\kappa), \label{toda:1}\\
\calD_q^\kappa b_n^{(\a)}(\kappa)&=&\frac{q}{1-q}\left[\left(\frac{a_n^{(\a)}(\kappa)}{1-\l \xi_{nn}^{(\a)}}\right)^2-\left(\frac{a_{n+1}^{(\a)}(\kappa)}{1-\l \xi_{n+1,n+1}^{(\a)}}\right)^2\right],\label{toda:2}
\eeqa
where $\xi_{nn}^{(\a)}(\kappa)$ is given in Eq.\eqref{3-1-b}.
\ethm
\bpf
We compute the $q$-derivative with respect to the $\kappa$ variable on the 
action of position operator, $x p_n^{(\a)}(x,\kappa)$, in two ways: 

We first compute the $q$-derivative, with respect to $\kappa$ of the recursive relation,
\beqa\label{ib}
\calD_q^\kappa[x p_n^{(\a)}(x,\kappa)]
&=&\calD_q^\kappa[a_{n+1}^{(\a)}(\kappa)p_{n+1}^{(\a)}(x,\kappa)+b_n^{(\a)}(\kappa)p_n^{(\a)}(x,\kappa)+a_n^{(\a)}(\kappa)p_{n-1}^{(\a)}(x,\kappa)]\nonumber\\
&=&[\calD_q^\kappa a_{n+1}^{(\a)}+\xi_{n+1,n+1}^{(\a)}\bar a_{n+1}^{(\a)}]p_{n+1}^{(\a)}+[\calD_q^\kappa b_n^{(\a)}+\xi_{n,n+1}^{(\a)} \bar a_{n+1}^{(\a)}+\xi_{nn}^{(\a)} \bar b_n^{(\a)}]p_n^{(\a)}\nonumber\\
&+&[\calD_q^\kappa a_n^{(\a)}+\xi_{n-1,n-1}^{(\a)} \bar a_n^{(\a)}+\xi_{n-1,n}^{(\a)}\bar b_n^{(\a)}]p_{n-1}^{(\a)}+[\xi_{n-2,n-1}^{(\a)} \bar a_n^{(\a)}]p_{n-2}^{(\a)}.
\eeqa
Here $\xi_{mn}^{(\a)}$ are the Fourier coefficients (matrix elements) of Eq.\eqref{ia}, and we suppress the dependence on $\kappa$ for 
simplicity. 

On the other hand, since $\calD_q^\kappa$ commutes with the position operator $x$, we first 
compute the $q$-derivative (w.r.t $\kappa$ variable) of the orthonormal polynomials and 
then apply the position operator.
\beqa\label{ic}
& &x \calD_q^\kappa[p_n^{(\a)}(x,\kappa)]\nonumber\\
&=&[\xi_{nn}^{(\a)} a_{n+1}^{(\a)}]p_{n+1}^{(\a)}+[\xi_{n-1,n}^{(\a)}  a_{n}^{(\a)}+\xi_{nn}^{(\a)} b_n^{(\a)}]p_n^{(\a)}\nonumber\\
&+&[\xi_{nn}^{(\a)} a_n^{(\a)}+\xi_{n-1,n}^{(\a)} b_{n-1}^{(\a)}]p_{n-1}^{(\a)}+[\xi_{n-1,n}^{(\a)} a_{n-1}^{(\a)}]p_{n-2}^{(\a)}.
\eeqa

By comparing the corresponding coefficients of each orthonormal polynomials as 
calculated in Eqs.\eqref{ib}, \eqref{ic}, we get the following set of relations:
\beqa
\calD_q^\kappa a_n^{(\a)}&=&\xi_{n-1,n-1}^{(\a)} a_n^{(\a)}-\xi_{nn}^{(\a)}\bar a_n^{(\a)},\label{id}\\
\calD_q^\kappa b_n^{(\a)}&=&\xi_{nn}^{(\a)}(b_n^{(\a)}-\bar b_n^{(\a)})+[\xi_{n-1,n}^{(\a)} a_n^{(\a)}-\xi_{n,n+1}^{(\a)} \bar a_{n+1}^{(\a)}],\\
\calD_q^\kappa a_n^{(\a)}&=&\xi_{n-1,n}^{(\a)}(b_{n-1}^{(\a)}-\bar b_n^{(\a)})+[\xi_{nn}^{(\a)} a_n^{(\a)}-\xi_{n-1,n-1}^{(\a)} \bar a_n^{(\a)}],\label{ie}\\
\xi_{n-2,n-1}^{(\a)}\bar a_n^{(\a)}&=&\xi_{n-1,n}^{(\a)}a_{n-1}^{(\a)}.
\eeqa
Note that the last result allows us to replace the rescaled recursive coefficients $\bar a_n^{(\a)}$
in terms of the Fourier coefficients and the unscaled recursive coefficients
\beq\label{5-1-2}
a_n^{(\a)}(q\kappa)=\frac{\xi_{n-1,n}^{(\a)}(\kappa)}{\xi_{n-2,n-1}^{(\a)}(\kappa)}a_{n-1}^{(\a)}(\kappa) 
=\frac{1-\l \xi_{n-1,n-1}^{(\a)}}{1-\l \xi_{nn}^{(\a)}}a_n^{(\a)}(\kappa),
\eeq
where the second equality of the above relation follows from Eq.\eqref{id}. We have also checked that Eqs.\eqref{id}, \eqref{ie} are compatible. 

Finally, after some manipulations, we 
get the coupled $q$-difference equations, Eqs.\eqref{toda:1}, \eqref{toda:2}.

\epf

\subsection{$q$-Difference equations for the recursive coefficients of the $q$-Hermite orthonormal polynomials}\
 \vspace{0.5cm}

In this section, we shall derive the $q$-difference equation for the recursive 
coefficients of the $q$-Hermite orthonormal polynomials. In order to achieve 
this, we need to compute the Fourier coefficients of the $q$-derivative
of the $q$-Hermite orthonormal polynomials with respect to parameter $\kappa$,
\beq\label{db}
\calD_q^\kappa P_n^{(\a)}(x;\kappa) = \sum_{j=0}^n P_j^{(\a)}(x;\kappa) \Xi_{jn}^{(\a)}(\kappa), \quad (n-j \mbox{ is even}).
\eeq
Note that, by recalling the definition of the $q$-derivative, Eq.\eqref{app:1}, we 
can also transform the equation above into the Fourier expansion of the 
$q$-evolved $q$-Hermite orthonormal polynomials with respect to parameter $\kappa$,
\beq\label{dc}
P_n^{(\a)}(x;q\kappa) =[1-\l \Xi_{nn}^{(\a)}]P_n^{(\a)}(x;\kappa)-\l \sum_{j=0}^{n-1}P_j^{(\a)}(x;\kappa)
\Xi_{jn}^{(\a)}(\kappa).
\eeq 

Due to the parity conserving property of the $q$-Hermite ensembles, we find that it is easier to firstly present the $\kappa$-deformation of the recursive coefficients in terms of $\Xi_{nn}^{(\a)}$ and $A_n^{(\a)}$.
\bthm
The $\kappa$-deformation of the recursive coefficients associated with the $q$-Hermite orthonormal polynomials is given by
\beq\label{toda:3}
\calD_q^\kappa A_n^{(\a)}(\kappa) =\frac{\Xi_{n-1,n-1}^{(\a)}(\kappa)-\Xi_{nn}^{(\a)}(\kappa)}{1+\kappa(q-1)\Xi_{nn}^{(\a)}(\kappa)}A_n^{(\a)}(\kappa).
\eeq
\ethm
\bpf
We compute 
the $\calD_q^\kappa$ derivative on the product $x P_n^{(\a)}(x;\kappa)$ in two ways.
\beqa\label{}
\calD_q^\kappa[x P_n^{(\a)}(x;\kappa)]&=&\calD_q^\kappa[A_{n+1}^{(\a)}(\kappa)P_{n+1}^{(\a)}(x;\kappa)+A_n^{(\a)}(\kappa)P_{n-1}^{(\a)}(x;\kappa)]\nonumber\\
&=&x [\sum_{j=0}^n P_j^{(\a)}(x;\kappa)\Xi_{jn}^{(\a)}(\kappa)]\nonumber\\
&=&\sum_{j=0}^n[A_{j+1}^{(\a)}(\kappa)P_{j+1}^{(\a)}(x;\kappa)+A_j^{(\a)}(\kappa)P_{j-1}^{(\a)}(x;\kappa)]\Xi_{jn}^{(\a)}(\kappa)].
\eeqa
By comparing the coefficients of $P_{n+1}^{(\a)}(x;\kappa)$ of the first and the third lines 
of the previous equation, we get the following results:
\beq\label{}
\calD_q^\kappa A_{n+1}^{(\a)}(\kappa)=  \Xi_{nn}^{(\a)}(\kappa) A_{n+1}^{(\a)}(\kappa)-\Xi_{n+1,n+1}^{(\a)}(\kappa) A_{n+1}^{(\a)}(q\kappa),
\eeq
which implies
\beq\label{5-2-3}
\frac{A_n^{(\a)}(q\kappa)}{1-\l \Xi_{n-1,n-1}^{(\a)}(\kappa)} =\frac{A_n^{(\a)}(\kappa)}{1-\l\Xi_{n,n}^{(\a)}(\kappa)},
\eeq
and
\beq\label{}
\calD_q^\kappa A_n^{(\a)}(\kappa) =\frac{\Xi_{n-1,n-1}^{(\a)}(\kappa)-\Xi_{nn}^{(\a)}(\kappa)}{1+\kappa(q-1)\Xi_{nn}^{(\a)}(\kappa)}A_n^{(\a)}(\kappa).
\eeq
\epf

Following similar discussion as Theorem \ref{sec2:3-5}, we can express the Fourier coefficients of the $\kappa$-deformation of the orthonormal polynomials in terms of the recursive coefficients $A_n^{(\a)}(\kappa)$.
\bthm
The Fourier coefficients of the $\kappa$-deformation of the orthonormal polynomials associated with the $q$-Hermite weight is given as $(\l:=(1-q)\kappa)$
\beq\label{}
\calD_q^\kappa P_n^{(\a)}(x,\kappa)=P_n^{(\a)}(x,\kappa) \Xi_{nn}^{(\a)}(\kappa)+P_{n-2}^{(\a)}(x,\kappa) \Xi_{n-2,n}^{(\a)}(\kappa),
\eeq
where,
 \beq\label{}
 \Xi_{n-2,n}^{(\a)}(\kappa)=\left(\frac{q^2 \kappa}{1-q}\right)\frac{ A_{n}^{(\a)} A_{n-1}^{(\a)}}{1-\l \Xi_{n,n}^{(\a)}},
 \eeq

 \beq\label{}
 \Xi_{n,n}^{(\a)}(\kappa)=\frac{1}{\sqrt{2}(1-q)\kappa}\left\{\sqrt{2}-\sqrt{1-\left(q\kappa\right)^2\bar A_n^{(\a)}(+)+\sqrt{1-2\left(q\kappa\right)^2\bar A_n^{(\a)}(+)+\left(q\kappa\right)^4\left[\bar B_n^{(\a)}\right]^2  }}\right\},
 \eeq
 \beq\label{}
\bar A_n^{(\a)}(+):=\left(\bar A_{n+1}^{(\a)}\right)^2+\left(\bar A_n^{(\a)}\right)^2, \mbox{ and }
\bar B_n^{(\a)}:=\left[\bar A_{n+1}^{(\a)}(+)\right]^2-4\left[ A_n^{(\a)}A_{n-1}^{(\a)}\right]^2.
 \eeq  
\ethm
\bpf
By taking the $q$-derivative (w.r.t $\kappa$) on the orthonormal 
condition
\beq\label{}
\calD_q^\kappa \left(\int_{-1}^1 P_m^{(\a)}(x;\kappa) P_n^{(\a)}(x;\kappa) \w^{(\a)}(x;\kappa) d_q x\right)=0,
\eeq
we get (assuming $m\leq n$)
\beqa\label{da}
& &\int_{-1}^1 \left(\calD_q^\kappa P_m^{(\a)}(x;\kappa)\right) P_n^{(\a)}(x;\kappa) \w^{(\a)}(x;\kappa) d_q x\nonumber\\
&+&\int_{-1}^1 P_m^{(\a)}(x;q\kappa) \left(\calD_q^\kappa P_n^{(\a)}(x;\kappa)\right) \w^{(\a)}(x;\kappa) d_q x\nonumber\\
&+&\int_{-1}^1 P_m^{(\a)}(x;q\kappa) P_n^{(\a)}(x;q\kappa) \left(\calD_q^\kappa \w^{(\a)}(x;\kappa)\right) d_q x=0.
\eeqa

Substituting the Fourier expansion, Eqs.\eqref{db}, into the first two terms of \eqref{da}, and 
using the Pearson relation (in the $\kappa$ variable) for the $q$-Hermite weight, we get 
\beqa
 & &\d_{mn} \Xi_{nn}^{(\a)}+(1-\l \Xi_{mn}^{(\a)})\Xi_{mn}^{(\a)} 
-\l \sum_{j=0}^{m-1} \Xi_{jm}^{(\a)}\Xi_{jn}^{(\a)}\nonumber\\
         &+&\d_{mn} \left(\frac{q^2\kappa}{q-1}\right)\left\{\left(\bar A_{n+1}^{(\a)}\right)^2+\left(\bar A_n^{(\a)}\right)^2\right\}
         +\d_{m,n-2}\left(\frac{q^2 \kappa}{q-1}\right) \bar A_{n-1}^{(\a)}\bar A_n^{(\a)}=0.
\eeqa

In order to illustrate the content of this equation, we consider the following specializations:\

\begin{itemize}
\item[Case $1$:] $m\leq n-3$\\
In this case, the master equation reduces to 
\beqs
(1-\l \Xi_{mm}^{(\a)}) \Xi_{mn}^{(\a)}-\l \sum_{j=0}^{m-1} \Xi_{j m}^{(\a)} \Xi_{j n}^{(\a)}=0.
\eeqs
By the mathematical induction, we show that $\Xi_{mn}^{(\a)}=0$, if $3\leq n-m$. Consequently, 
the Fourier expansion of the $q$-derivative (w.r.t $\kappa$ variable) on the $q$-Hermite 
orthonormal polynomials only consist of two terms:
\beq\label{}
\calD_q^\kappa P_n^{(\a)}(x,\kappa)=P_n^{(\a)}(x,\kappa) \Xi_{nn}^{(\a)}(\kappa)+P_{n-2}^{(\a)}(x,\kappa) \Xi_{n-2,n}^{(\a)}(\kappa).
\eeq

\item[Case $2$:] $m=n-2$\\
In this case, the master equation reduces to 
\beqs
(1-\l \Xi_{n-2,n-2}^{(\a)}) \Xi_{n-2,n}^{(\a)}-\l \sum_{j=0}^{n-3} \Xi_{j,n-2}^{(\a)} \Xi_{j n}^{(\a)}+\frac{q^2 \kappa}{q-1}\bar A_n^{(\a)}\bar A_{n-1}^{(\a)}=0.
\eeqs
Since we have showed that $\Xi_{j n}^{(\a)}=0$ for $j\leq n-3$ in Case $1$, we can use the 
 equation above to express the off-diagonal Fourier coefficient in terms of the diagonal ones:
 \beq\label{5-2-1}
 \Xi_{n-1,n+1}^{(\a)}=\left(\frac{q^2 \kappa}{1-q}\right)\frac{\bar A_{n+1}^{(\a)} \bar A_n^{(\a)}}{1-\l \Xi_{n-1,n-1}^{(\a)}}=\left(\frac{q^2 \kappa}{1-q}\right)\frac{ A_{n+1}^{(\a)} A_n^{(\a)}}{1-\l \Xi_{n+1,n+1}^{(\a)}}. 
 \eeq
 For the second equality of the equation above, we use Eq.\eqref{5-2-3} to replace $\bar A_n^{(\a)}$ in terms of $A_n^{(\a)}$.
 
 \item[Case $3$:] $m=n-1$\\
 Due to the parity preserving property associated with the $q$-Hermite ensemble,\\ $\Xi_{n-1,n}^{(\a)}=0$, we 
 have no constraint in this case.\\
 
 \item[Case $4$:] $m=n$\\
 In this case, we have 
 \beq\label{}
 \Xi_{nn}^{(\a)}+(1-\l \Xi_{nn}^{(\a)})\Xi_{nn}^{(\a)}- \l(\Xi_{n-2,n}^{(\a)})^2=\left(\frac{q^2 \kappa}{1-q}\right)\left[\left(\bar A_{n+1}^{(\a)}\right)^2+\left(\bar A_n^{(\a)}\right)^2\right].
 \eeq  
 After suitable rearrangement, we get
 \beq\label{5-2-2}
 (1-\l \Xi_{nn}^{(\a)})^2+\frac{(q\kappa)^4\left(\bar A_n^{(\a)}\right)^2\left(\bar A_{n-1}^{(\a)}\right)^2}{ (1-\l \Xi_{n-2,n-2}^{(\a)})^2}
 =1-(q\kappa)^2\left\{\left(\bar A_{n+1}^{(\a)}\right)^2+\left(\bar A_n^{(\a)}\right)^2\right\}.
 \eeq
\end{itemize}

On the other hand, by replacing $\bar A_n^{(\a)}$ by $A_n^{(\a)}$, using Eq.\eqref{5-2-1}, we derive a quadratic equation for $(1-\l \Xi_{nn}^{(\a)})^2$,
 \beq\label{5-2-4}
 (1-\l \Xi_{nn}^{(\a)})^2+\frac{(q\kappa)^4\left(A_n^{(\a)}\right)^2\left( A_{n-1}^{(\a)}\right)^2}{ (1-\l \Xi_{nn}^{(\a)})^2}
 =1-(q\kappa)^2\left\{\left(\bar A_{n+1}^{(\a)}\right)^2+\left(\bar A_n^{(\a)}\right)^2\right\}.
 \eeq
Hence, $\Xi_{nn}^{(\a)}$ can be solved in terms of $A_n^{(\a)}$ and $\bar A_n^{(\a)}$.
\epf

By substituting the solution of $\Xi_{nn}^{(\a)}$ for Eq.\eqref{5-2-4} back to Eq.\eqref{toda:3}, we get closed $q$-difference equations for the 
recursive coefficients of the generalized $q$-Hermite ensemble.

\subsection{On the compatibility of the quadratic relation and $q$-Toda equations}\
 \vspace{0.5cm}

In this section, we check the compatibility between the quadratic relation 
Eqs. \eqref{bc}, \eqref{bd},\eqref{be}, \eqref{bf} and the $q$-Toda equation Eqs. \eqref{toda:1}, \eqref{toda:2}, \eqref{toda:3}. To see this, we first example the 
Fourier coefficients of the $q$-derivative of the orthonormal $q$-Laguerre/Hermite 
polynomials (w.r.t $\kappa$). 

\bthm
The Fourier coefficients of the $q$-derivative (w.r.t $\kappa$) of the orthonormal 
$q$-Laguerre/Hermite polynomials (Eqs. \eqref{ia}, \eqref{db}) are related by the quadratic 
relations
\beqa
\Xi_{2n,2n}^{(\a)}(\kappa,q)&=&(1+q)\kappa\xi_{nn}^{(\a)}(\kappa^2,q^2),\nonumber\\
\Xi_{2n+1,2n+1}^{(\a)}(\kappa,q)&=&(1+q)\kappa\xi_{nn}^{(\a+1)}(\kappa^2,q^2).\nonumber
\eeqa
\ethm
\bpf
We relate the $q$-derivative (w.r.t $\kappa$) of the orthonormal $q$-Hermite 
polynomials Eq. \eqref{bf} to that of the $q$-Laguerre polynomials in two ways. 
First of all, for even polynomials
\begin{align}\label{5-3-1}
&\calD_q^\kappa P_{2n}^{(\a)}(x;\kappa,q)\nonumber\\
&=\frac{P_{2n}^{(\a)}(x;\kappa,q)-P_{2n}^{(\a)}(x;q\kappa,q)}{(1-q)\kappa}\nonumber\\
&=\frac{p_{n}^{(\a)}(x^2;\kappa^2,q^2)-p_{n}^{(\a)}(x^2;q^2\kappa^2,q^2)}{(1-q^2)\kappa^2}\frac{(1+q)^{\frac{3}{2}}\kappa}{\sqrt{2}}\nonumber\\
&=\frac{(1+q)^{\frac{3}{2}}\kappa}{\sqrt{2}}\left[\calD_{q^2}^{\kappa^2}p_{n}^{(\a)}(x^2;\kappa^2,q^2)\right]\nonumber\\
&=\frac{(1+q)^{\frac{3}{2}}\kappa}{\sqrt{2}}\left[p_{n}^{(\a)}(x^2;\kappa^2,q^2)\xi_{nn}^{(\a)}(\kappa^2,q^2)+p_{n-1}^{(\a)}(x^2;\kappa^2,q^2)\xi_{n-1,n}^{(\a)}(\kappa^2,q^2)\right].
\end{align}
On the other hand, if we write the Fourier expansion of the $q$-derivative (w.r.t $\kappa$) of the $q$-Hermite 
polynomials
\begin{align}\label{5-3-2}
&\calD_q^\kappa P_{2n}^{(\a)}(x;\kappa,q)\nonumber\\
&= P_{2n}^{(\a)}(x;\kappa,q) \Xi_{2n,2n}^{(\a)}(\kappa,q)+P_{2n-2}^{(\a)}(x;\kappa,q)\Xi_{2n-2,2n}^{(\a)}(\kappa,q)\nonumber\\
&=\sqrt{\frac{1+q}{2}}p_{n}^{(\a)}(x^2;\kappa^2,q^2)\Xi_{2n,2n}^{(\a)}(\kappa,q)+\sqrt{\frac{1+q}{2}}p_{n-1}^{(\a)}(x^2;\kappa^2,q^2)\Xi_{2n-2,2n}^{(\a)}(\kappa,q).
\end{align}
By comparing the two results Eqs. \eqref{5-3-1}, \eqref{5-3-2}, we obtain
\beqa
\Xi_{2n,2n}^{(\a)}(\kappa,q)&=&(1+q)\kappa \xi_{nn}^{(\a)}(\kappa^2,q^2),\label{5-3-5}\\
\Xi_{2n-2,2n}^{(\a)}(\kappa,q)&=&(1+q)\kappa \xi_{n-1,n}^{(\a)}(\kappa^2,q^2).\label{5-3-6}
\eeqa
Next, we compare the odd $q$-Hermite polynomials.
\begin{align}\label{5-3-3}
&\calD_q^\kappa P_{2n+1}^{(\a)}(x;\kappa,q)\nonumber\\
&=\frac{(1+q)^{\frac{3}{2}}\kappa}{\sqrt{2}} x \calD_{q^2}^{\kappa^2}p_{n}^{(\a+1)}(x^2;\kappa^2,q^2) \nonumber\\
&=\frac{(1+q)^{\frac{3}{2}}\kappa}{\sqrt{2}} x \left[p_{n}^{(\a+1)}(x^2;\kappa^2,q^2)\xi_{nn}^{(\a+1)}(\kappa^2,q^2)+p_{n-1}^{(\a+1)}(x^2;\kappa^2,q^2)\xi_{n-1,n}^{(\a+1)}(\kappa^2,q^2)\right],
\end{align}
and
\begin{align}\label{5-3-4}
&\calD_q^\kappa P_{2n+1}^{(\a)}(x;\kappa,q)\nonumber\\
&= P_{2n+1}^{(\a)}(x;\kappa,q) \Xi_{2n+1,2n+1}^{(\a)}(\kappa,q)+P_{2n-1}^{(\a)}(x;\kappa,q)\Xi_{2n-1,2n+1}^{(\a)}(\kappa,q)\nonumber\\
&=\sqrt{\frac{1+q}{2}}x p_{n}^{(\a+1)}(x^2;\kappa^2,q^2)\Xi_{2n+1,2n+1}^{(\a)}(\kappa,q)+\sqrt{\frac{1+q}{2}}x p_{n-1}^{(\a+1)}(x^2;\kappa^2,q^2)\Xi_{2n-1,2n+1}^{(\a)}(\kappa,q).
\end{align}
By comparing the two results, Eqs. \eqref{5-3-3}, \eqref{5-3-4}, we obtain
\beqa
\Xi_{2n+1,2n+1}^{(\a)}(\kappa,q)&=&(1+q)\kappa \xi_{nn}^{(\a+1)}(\kappa^2,q^2),\label{5-3-7}\\
\Xi_{2n-1,2n+1}^{(\a)}(\kappa,q)&=&(1+q)\kappa \xi_{n-1,n}^{(\a+1)}(\kappa^2,q^2).\label{5-3-8}
\eeqa
\epf

In fact, the quadratic relation among the Fourier coefficients of the $q$-Laguerre/Hermite 
polynomials are equivalent to the quadratic relation among the recursive coefficients of 
the $q$-Laguerre/Hermite polynomials. To see this, we rewrite the Eq. \eqref{5-2-1} (set $n=m-1$) as
\beq\label{}
A_m^{(\a)}(\kappa,q)A_{m-1}^{(\a)}(\kappa,q)=\frac{1-q}{q^2\kappa}\Xi_{m-2,m}^{(\a)}(\kappa,q)\left[1-(1-q)\kappa \Xi_{mm}^{(\a)}(\kappa,q)\right].
\eeq
For $m=2n$, after substituting Eqs. \eqref{5-3-5}, \eqref{5-3-6} and using Eq. \eqref{5-1-1}, we get
\beq\label{}
A_{2n}^{(\a)}(\kappa,q)A_{2n-1}^{(\a)}(\kappa,q)=\frac{1-q^2}{q^2}\xi_{n-1,n}^{(\a)}(\kappa^2,q^2)\left[1-(1-q^2)\kappa^2 \xi_{nn}^{(\a)}(\kappa^2,q^2)\right]=a_n^{(\a)}(\kappa^2,q^2).
\eeq 
For $m=2n+1$, after substituting Eqs. \eqref{5-3-7} , \eqref{5-3-8} and using Eq. \eqref{5-1-1}, we get
\beq\label{}
A_{2n+1}^{(\a)}(\kappa,q)A_{2n}^{(\a)}(\kappa,q)=\frac{1-q^2}{q^2}\xi_{n-1,n}^{(\a+1)}(\kappa^2,q^2)\left[1-(1-q^2)\kappa^2 \xi_{nn}^{(\a+1)}(\kappa^2,q^2)\right]=a_n^{(\a+1)}(\kappa^2,q^2).
\eeq 

Similarly, rewriting Eq. \eqref{5-2-2},
\begin{align}
&(q\kappa)^2\left[\left(A_{m+1}^{(\a)}(q\kappa,q)\right)^2+\left(A_{m}^{(\a)}(q\kappa,q)\right)^2\right]\nonumber\\
&=1-[1-(1-q)\kappa \Xi_{mm}(\kappa,q)]^2-\frac{(q\kappa)^4\left[A_{m}^{(\a)}(\kappa,q)\right]^2\left[A_{m-1}^{(\a)}(\kappa,q)\right]^2}{[1-(1-q)\kappa \Xi_{mm}(\kappa,q)]^2},
\end{align}
then by substituting Eqs.\eqref{bc}, \eqref{be}, \eqref{if}, \eqref{5-3-5}, \eqref{5-3-6}, for either $m=2n$ or $m=2n+1$, we 
reproduce Eqs.\eqref{bd}, \eqref{bf}.

\section{Summary and Conclusion}

In this paper, we study the exponential deformation/evolution of the generalized little $q$-Laguerre/Hermite polynomials. Our study serves as a $q$-discrete generalization for the correspondence between the Lax equation of the Toda lattice and the exponential deformation of any orthogonal polynomial system. In addition, we also discuss the implications and compatibility of the quadratic relation among $q$-Laguerre and $q$-Hermite orthogonal system. While it is not clear, at the present stage if we can write down a $q$-discrete Lax equation for the systems under study, but our calculation at least provide a possible hint of further explorations. Furthermore, it is of interest to compare with other approaches \cite{33C45} for this problem, to see if there are simpler expressions for the results obtained in this paper.  

\section*{Acknowledgement}

This research project was initiated in a summer visit (by C. T.) to the Institute of Mathematics 
at Academia Sinica in 2015. Both authors would like to thank Derchyi Wu, Chueh-Hsin Chang, and Mourad E. H. Ismail for instructive discussions. C.T. 
would like to thank Derchyi Wu for invitation and hospitality. The research work 
of C.T. is partially supported by the grant from Academia Sinica for summer visit,
and in parts supported by the MOST research grant 104-2112-M-029-001. The 
research work of H.F. is partially supported by the MOST research 
grants 104-2115-M-001-001-MY2, 105-2112-M-29-003 and 106-2112-M-29-005.

\newpage

\appendix
\section{Some Basic Definitions and Relations for 
              $q$-Analysis ($0<q<1$)}\label{app:a}



In this section, we collect some basic definitions and formulas which 
are relevant to our discussions.

The $q$-integral (Jackson integral) for a function $f(x)$ over the region $x\in [0,a]$ is defined as
\beqa
F(a)&:=&\int_0^a f(x) d_q x := a(1-q)\sum_{k=0}^\infty f(a q^k)q^k,\label{app:3}\\
F(a,b)&:=&\int_b^a f(x) d_q x:=F(a)-F(b).
\eeqa
This is compatible with the definition of the $q$-derivative
\beq\label{app:1}
\calD_q^x f(x) := \frac{f(q x)-f(x)}{qx-x}=\frac{f(x)-f(q x)}{x(1-q)}
\eeq
in the following senses:

\ben
\item Fundamental theorem of the $q$-calculus
\beqa
\calD_q^a F(a)&=&f(a),\\
\int_0^a [\calD_q^x f(x)] d_q x&=&f(a)-f(0).
\eeqa

\item The linear change of variables can be implemented in $q$-integral:
\beq\label{}
\int_0^a f(c x) d_q x=\frac{1}{c}\int_0^{ca} f(y) d_q y.
\eeq
\een

There are some subtleties associated with the $q$-derivative, in particular, the 
$q$-Lebinitz rule is given as
\beqa\label{app:2}
\calD_q^x[f(x)g(x)] & = & \frac{f(qx)g(qx)-f(x)g(x)}{(q-1)x} \nonumber\\
                           & = & f(qx)[\calD_q^x g(x)]+[\calD_q^x f(x)]g(x) \nonumber\\
                           & = & [\calD_q^x f(x)]g(qx)+f(x)[\calD_q^x g(x)].
\eeqa

\bibliographystyle{plain}

\end{document}